\documentclass[10pt,preprint]{aastex}
 \usepackage{natbib}
\def\etal{et al.~}

\def\ang{\AA}

\def\gapprox{\lower.4ex\hbox{$\;\buildrel >\over{\scriptstyle\sim}\;$}}
\def\lapprox{\lower.4ex\hbox{$\;\buildrel <\over{\scriptstyle\sim}\;$}}

\shortauthors{Aschwanden 2022}
\shorttitle{Power law slopes in SOC systems}

\begin{document}
\renewcommand{\topfraction}{0.95}
\renewcommand{\bottomfraction}{0.95}
\renewcommand{\textfraction}{0.05}
\renewcommand{\floatpagefraction}{0.95}
\renewcommand{\dbltopfraction}{0.95}
\renewcommand{\dblfloatpagefraction}{0.95}


\title{Reconciling Power Law Slopes in Solar Flare and Nanoflare 
       Size Distributions }

\author{Markus J. Aschwanden}
\affil{Lockheed Martin, Solar and Astrophysics Laboratory (LMSAL),
       Advanced Technology Center (ATC),
       A021S, Bldg.252, 3251 Hanover St.,
       Palo Alto, CA 94304, USA;
       e-mail: aschwanden@lmsal.com}

\begin{abstract}
We unify the power laws of size distributions of solar flare 
and nanoflare energies. We present three models that predict the
power law slopes $\alpha_E$ of flare energies
defined in terms of the 2-D and 3-D 
fractal dimensions ($D_A, D_V$): (i) The spatio-temporal
standard SOC model, defined by the power law slope 
$\alpha_{E1}=1+2/(D_V+2)=(13/9)\approx 1.44$; 
(ii) the 2-D thermal energy model,
$\alpha_{E2}=1+2/D_A=(7/3)\approx 2.33$, and
(iii) the 3-D thermal energy model, 
$\alpha_{E3}=1+2/D_V=(9/5)\approx 1.80$. 
The theoretical predictions of energies
are consistent with the observational values 
of these three groups, i.e., 
$\alpha_{E1}=1.47 \pm 0.07$; 
$\alpha_{E2}=2.38 \pm 0.09$, and
$\alpha_{E3}=1.80 \pm 0.18$. 
These results corroborate that the energy of nanoflares
does not diverge at small energies, since $(\alpha_{E1}<2$)
and $(\alpha_{E3}<2)$, except for the unphyiscal 2-D model
$(\alpha_{E2}>2)$. This conclusion adds an additional
argument against the scenario of coronal heating by nanoflares.
\end{abstract}


\section{	Introduction 				}  

The concept of {\sl self-organized criticality (SOC)} 
specifies nonlinear (avalanching) phenomena based on next-neighbor 
interactions in a lattice grid (Bak et al.~1987; 1988).
The spatio-temporal evolution of avalanches in such complex 
environments has been numerically simulated by cellular 
automaton methods, which exhibit fractal structures (Pruessner 2012).
Alternatively, SOC-related avalanches can be considered
analytically, as instabilities with a nonlinear initial 
growth phase and subsequent saturation (Rosner et al.~1978; 
Aschwanden 2011, 2012, 2014). 

Let us introduce the first definition of a SOC energy,
which we call the {\sl spatio-temporal energy} of a SOC 
avalanche, labeled as $E_1$ in this Letter. 
The analytical SOC approach
takes both the spatial as well as the temporal evolution
of a SOC avalanche into account. Hence, the total energy 
of a SOC avalanche is determined from the spatial and 
temporal integration over the unstable pixel areas
(or voxel volumes). 
Theoretical predictions of SOC parameters based on the
size distribution of avalanche energies are described in
Section 2 and are summarized in Table 1. 
The necessary parameters to characterize the spatio-temporal
energy requires the measurement of the mean flux $F$
and the time duration $T$ of an event,
while the spatio-temporal energy is defined by $E_1 = F * T$.
Measurements of spatio-temporal energies were originally
applied to soft X-rays in solar flares (Drake 1971),
hard X-rays (Crosby et al.~1993;
Lu et al.~1993), gamma rays (Perez-Enriquez and 
Miroshnichenko 1999), as well as to EUV small-scale
brightenings or nanoflares (Brkovic et al.~2001;
Uritsky et al.~2013). We compile these data sets in
Table 2. 

The second definition of a SOC energy was introduced
by the 2-D definition of the thermal energy
in a high-temperature plasma
(called $E_2$ here), i.e., $E_2=3 k_B T_e n_e V$
(where $k_B$ is the Boltzmann constant, $T_e$ the electron
temperature, $n_e$ the electron density, and $V$ the
fractal volume. The avalanche volumes of solar flares
and nanoflares are generally close to fractal geometries, 
if they are measured on a pixel-by-pixel basis 
(Aschwanden and Aschwanden 2008a, 2008b), in contrast to
an encompassing (non-fractal) circle or square 
(with length scale $L$). An additional relationship 
is the definition of the emission measure,
$EM = n_e^2 \ V$, which can be used to substitute the
electron density, i.e., $n_e = \sqrt{EM / V}$. 
Furthermore, a relationship for the (fractal) event
volume $V$ needs to be specified. The projected area
$A$, which is fractal, can be measured directly in
the image plane, but the line-of-sight depth is
unknown, which led some pioneers to quantify it
with a constant height $h_0$, leading to the
expression $V = A\ h_0$ for the volume 
(Krucker and Benz 1998; Parnell and Jupp 2000;
Benz and Krucker 2002, Table 3). In hindsight, this choice
of a constant depth $h_0$ has been criticized to be 
unphysical, because it is very unlikely that the
line-of-sight depth is equal from the smallest
nanoflares (of the size of a solar granulation cell)
to the largest flares (of the size of an active region). 
Furthermore, the assumption
of a constant height $h_0$ introduces a crucial
bias that modifies the power law slope of the energy
size distribution substantially, from 
$\alpha_{E3}=1.80$ to $\alpha_{E2}=2.33$, 
across the critical value of $\alpha_E=2$,
as we will see in the remainder of this Letter. 

A third definition of a SOC energy ($E_3$) was made by 
abandoning the unphysical assumption of a
constant height, i.e., $V = A \ h_0$, and
instead replacing it with a more physical
assumption of an isotropic volume, i.e., 
$V = A^{3/2}$, which corresponds to a
line-of-sight depth of $h=\sqrt{A}$,
while $D_A$ is the fractal dimension of
the area, $A = L^{D_A}$. This approach,
which we call the 3-D thermal energy model
(Table 4), has been applied frequently 
(Shimizu 1995; Berghmans et al.~1998;
Berghmans and Clette 1999; Parnell and Jupp 2000;
Aschwanden and Parnell 2002; Uritsky et al.~2007;
Joulin et al.~2016; Nhalil et al.~2020;
Purkhart and Veronig 2022; Kawai and Imada 2022).

In this Letter we calculate the power law indices
$\alpha_E$ of the energy size distributions 
in a unified way, which
provides us a diagnostic whether flare and nanoflare
energies diverge at the lower or upper end of the
size distribution, and this way we can assess the importance
(or non-importance)
of coronal heating by nanoflares. The mathematical
derivation of the SOC models is given in Section 2,
a discussion in Section 3, and conclusions in
Section 4. 

\section{	Theoretical Models and Observations 	}

A SOC model should be able to predict the (occurrence frequency)
size distribution functions, which can be formulated in terms of
power law function slopes $\alpha_x$ to first order. Common SOC
parameters $x=[L,A,V,T,F,P,E]$ include the length scale $L$,
the 2-D area $A$, the 3-D volume $V$, the time duration $T$,
the mean flux or intensity $F$, the peak flux or peak intensity $P$,
and the fluence or energy $E$. In this study we reconcile three 
different definitions of the energy ($E$) that have been used in the past, 
namely the spatio-temporal definition of the standard SOC model ($E_1$), 
the 2-D fractal model ($E_2$), and the 3-D fractal model ($E_3$).

\subsection{		The Standard SOC Model 			}

The standard SOC model is derived from first principles in previous
studies (Aschwanden 2012, 2014, 2022). A brief summary of the
calculations that clarify the assumptions made here is given
in the following, while a more detailed description is provided
in Aschwanden (2022).

We start with the size distribution $N(L)$ of length scales $L$, 
also called the {\sl scale-free probability conjecture},
\begin{equation}
	N(L) \ dL \propto L^{-d} \ dL \ ,
\end{equation}
where $d$ is the Euclidean space dimension, which is set to $d=3$ 
for most real-world data. The power law indices $\alpha_x$ of the size
distribution functions can then be calculated for every SOC
parameter $x$ by variable substitution $L \mapsto x$,
\begin{equation}
	N(x) \ dx = L(x)^{-d} \ {dL \over dx} \ dx = x^{-\alpha_x} \ dx  \ ,
\end{equation}
which just requires the scaling law $x(L)$ as a function of
the length scale $L$, the inverted
scaling law function $L(x)$, and its derivative $dL/dx$.
Thus, we need to make an assumption of a scaling law
$x(L)$ of each SOC parameter of interest. 

For the spatial parameters
we define the fractal dimensions for 2-D areas $A$, which is
identical to the fractal Hausdorff dimension $D_A \approx 3/2$,
\begin{equation}
	A = L^{D_A} \ ,
\end{equation}
and like-wise for the 3-D volume $V$, which is 
identical to the fractal Hausdorff dimension $D_V \approx 5/2$,
\begin{equation}
	V = L^{D_V} \ .
\end{equation}
We can estimate the numerical values of the fractal dimensions 
$D_A$ and $D_V$ from the mean of the minimum and maximum value
in each Euclidean domain,
\begin{equation}
	D_A = {(D_{A,min} + D_{A,max}) \over 2} = {(1 + 2)\over 2} = {3 \over 2} = 1.50 \ ,
\end{equation} 
and correspondingly, 
\begin{equation}
	D_V = {(D_{V,min} + D_{V,max}) \over 2} = {(2 + 3)\over 2} = {5 \over 2} = 2.50 \ .
\end{equation} 
In the standard SOC model we need four more scaling laws. The time duration
$T$ of a SOC avalanche can be linked to spatial (fractal) structures by 
the diffusive behaviour,
\begin{equation}
	T \propto L^{2/\beta} \ ,
\end{equation}
where the coefficient is $\beta=1$ for classical diffusion,
$\beta < 1$ for sub-diffusive transport, and
$\beta > 1$ for hyper-diffusive transport (also called Levy flight).
Furthermore we need a relationship between the mean flux $F$ and the
emitting volume $V$,
\begin{equation}
	F \propto V^{\gamma} = L^{D_V \gamma} \ ,
\end{equation}
which is generally found to be near to proportional, hence we set $\gamma=1$.
A relationship between the peak flux $P$ and the length scale $L$ is,
\begin{equation}
	P \propto V^{\gamma} = L^{d \gamma} \ ,
\end{equation}
where the flux $F$ (Eq.~8) is maximized to the peak flux $P$, i.e.,
$P(t_{peak})=max[F(t)]$, by replacing the dimension $D_V$ in Eq.~(8)
with the Euclidean dimension $d$, i.e., $D_V \mapsto d$.

Finally, the fluence or energy $E_1$, which is expressed by the product
of the mean flux $F$ and the event duration $T$ (for a spatio-temporal
SOC event) yields (Crosby et al.~1993),
\begin{equation}
	E_1	= (F * T) \propto L^{(D_V \gamma + 2/ \beta)} \ .
\end{equation}
If we assume classical diffusion ($\beta=1$) and flux-volume proportionality
($\gamma=1$), the four basic scaling laws are reduced further to
$T \propto L^2$ (Eq.~7),
$F \propto L^{2.5}$ (Eq.~8),
$P \propto L^{3}$ (Eq.~9), and
$E_1\propto L^{4.5}$ (Eq.~10).
In this framework, there are no free parameters, and the power law slopes 
$\alpha_x$ of the size distributions,
\begin{equation}
	N(x)\ dx = x^{-\alpha_x} \ dx \ , 
\end{equation}
of all SOC parameters $x=[A,V,T,F,P,E]$ can be predicted by variable
substitution (Eq.~2), yielding the values 
$D_A=3/2$, 
$D_V=5/2$, 
$\alpha_A=7/3 \approx 2.33$, 
$\alpha_V=9/5 \approx 1.80$, 
$\alpha_T=2$, 
$\alpha_F=9/5 \approx 1.80$, 
$\alpha_P=5/3 \approx 1.67$, and 
$\alpha_{E_1}=13/9\approx 1.44$, as listed in Table 1. 

Comparison of these theoretical predictions of power law slopes
$\alpha_x^{theo}$ with observed size distributions $\alpha_x^{obs}$
have been presented in Aschwanden (2022). Among the solar flare data sets
that apply the spatio-temporal energy model (Eq.~10;
Table 2), we identify hard X-ray data (Crosby et al.~1993;
Lu et al.~1993), gamma ray data (Perez-Enriquez and Miroshnichenko 1999),
soft X-ray data (Drake 1971), and EUV data 
(Brkovic et al.~2001; Uritsky et al.~2013),
which exhibit a mean power slope of $\alpha_{E_1}^{obs}=1.47\pm 0.07$,
agreeing well with the theoretical prediction 
$\alpha_{E_1}^{theo}=(13/9)\approx 1.44$.

\subsection{		The 2-D Thermal Energy Model 		}

The energy of a spatio-temporal SOC event is defined in the standard
SOC model by the product of the count rate $(F)$ and the event duration $(T)$
(Eq.~10), which is appropriate for nonthermal energies that are quantified
by hard X-ray counts (or intensity) in solar and stellar flares. In both
solar or stellar flares, down to nanoflares, one can estimate thermal
(radiative) energies at the peak time of an event, defined by
\begin{equation}
	E_2 = ( 3 k_B n_e T_e ) \ V \ ,
\end{equation}
where $k_B$ is the Boltzmann constant, $n_e$ the electron density,
$T_e$ the electron temperature, and $V$ the 3-D volume, all measured
at the peak time of an event. The 3-D volume $V$
cannot be measured directly, which led some authors to approximate the
volume with a constant height $h_0$ in the line-of-sight,
\begin{equation}
	V = A \ h_0 = L^{D_A} \ h_0 \ ,
\end{equation}
while the fractal area is defined as $A=L^{D_A}$. 
The fractality is not explicitly mentioned in some of these studies,
but every pattern recognition code that measures an area on a pixel-by-pixel
basis (at different spatial resolutions) 
yields approximately the fractal area $A \propto L^{D_A}$ with
$D_A < 2$, rather than the encompassing Euclidean area $A=L^2$.
Inserting the area fractal dimension $D_A=3/2$ (Eq.~5) into the expression
for the thermal energy $E_2 \propto L^{D_A}$ (Eq.~12), we obtain 
\begin{equation}
	E_2 = ( 3 k_B n_e T_e h_0 ) \ L^{D_A} \ .
\end{equation}
The same way as we substituted the variable $L$ in the size distribution
with the energy $x=E$ (Eqs.~1 and 2),
\begin{equation}
	N(E_2) \ dE_2 = L(E_2)^{-d} \ {dL \over dE_2} \ dE_2 
	= E_2^{-\alpha_{E_2}} \ dE_2 \ ,
\end{equation}
yielding the power law slope $\alpha_{E_2}$, for $d=3$ and $D_A=3/2$, 
\begin{equation}
	\alpha_{E_2} = 1 + {(d - 1) \over D_A} 
		    = {7 \over 3} \approx 2.33 \ .
\end{equation}
Note that we treat the variables $n_e$, $T_e$, $h_0$ 
as constants here, while the scaling law hinges entirely on the 
correlation between the thermal energy $E_2$ and the length scale $L$, 
rendering a first-order approximation to the power law slope
$\alpha_{E_2}$. Since the thermal energy $E_2 \propto V \propto L^{D_A}$
(Eq.~14) and the fractal area $A \propto L^{D_A}$ (Eq.~13) have
the same scaling law, the power law index for the size distribution 
of areas $\alpha_A$ has the same power law index $\alpha_{E_2}$ too,
\begin{equation}
	\alpha_A = \alpha_{E_2} = {7 \over 3} \approx 2.33 \ .
\end{equation}
The definition of the energy made here (Eq.~12) invokes an isothermal
plasma. Nevertheless, the definition of the thermal energy can accomodate
a multi-thermal formalism, which involves a differential
emission measure distribution function $dEM(T_e)/dT_e$, characterized by
the increase in the emission measure $EM$, the (mean) electron density $n_e$,
and the volume $V$,
\begin{equation}
	EM = n_e^2 \ V \ ,
\end{equation}
which inserted into Eq.~(12) yields,
\begin{equation}
	E_2 = ( 3 k_B T_e ) \ \sqrt{ EM * V      }
	   =  ( 3 k_B T_e ) \ \sqrt{ EM * A \ h_0} \ .
\end{equation}
Size distribution of thermal energies, based on emission measure changes
$EM$, yield power law slopes of $\alpha_{E_2}=2.38\pm0.09$
(Krucker and Benz 1998; Parnell and Jupp 2000; Benz and Krucker 2002),
which match closely the theoretically expected value of
$\alpha_E=2.33$ (Eq.~16).

\subsection{		The 3-D Thermal Energy Model 		}

In the 3-D version of the thermal model, the SOC avalanche volume
$V \propto L^{D_V}$ (Eq.~4) is defined by the (mean) Hausdorff 
dimension $D_V=(5/2)$ (Eq.~6), which inserted into the expression
for the thermal energy is, 
\begin{equation}
	E_3 = ( 3 k_B n_e T_e ) \ L^{D_V} \ .
\end{equation}
We substitute the variable $L$ in the size distribution
of the thermal energy $E_3$ (Eq.~14),
\begin{equation}
	N(E_3) \ dE_3 = L(E_3)^{-d} \ {dL \over dE_3} \ dE_3 = E_3^{-\alpha_{E_3}} \ dE_3 \ ,
\end{equation}
yielding the power law slope $\alpha_{E_3}$, for $d=3$ and $D_V=5/2$, 
\begin{equation}
	\alpha_{E_3} = 1 + {(d - 1) \over D_V} 
		    = {9 \over 5} = 1.80 \ .
\end{equation}
Note that the power law slope is substantially steeper in the 
2-D model ($\alpha_{E_2}=2.33$) than in the 3-D version ($\alpha_{E_3}=1.80$).
Moreover, the two models predict power law slopes below ($\alpha_{E_2}<2$),
as well as above ($\alpha_{E_3}>2$) the critical value of $\alpha_E=2$,
which decides whether the nanoflare population diverges at the low end
or upper end of the size distribution.
Calculations of the multi-thermal energy using a 3-D model have been
performed using Yohkoh, TRACE, SOHO, AIA, and IRIS data
(Shimizu 1995; Berghmans et al.~1998; Berghmans and Clette 1999;
Parnell and Jupp 2000; Aschwanden and Parnell 2002; Uritsky et al.~2007;
Joulin et al.~2016; Nhalil et al.~2020; Purkhart and Veronig 2022; 
Kawai and Imada 2022), as listed in Table 4.

\section{	Discussion		}

\subsection{	Scaling Laws 		}

Scaling laws, typically expressed by variables $(x,y,...)$ with
power law dependencies, $x^{\alpha} y^{\beta}$ ... = const, are 
powerful tools to test parameter correlations and size distribution 
functions. If a scaling law function $y(x)$ and a single
size distribution $N(x)$ is known, we can derive the size distribution
$N(y)$ of a correlated parameter by variable substitution, 
$N(y) dy = N(x[y]) (dx/dy) dy$. This way we can predict theoretical size
distributions $N(y)$ based on observed size distributions $N(x)$,
as well as significant correlations between variables.
Here we explore the size distributions of 9 variables 
$x=[L, A, V, T, F, P, E_1, E_2, E_3]$ in a unified scheme (Table 1). 
We focus mainly on the three energy parameters
$x=[E_1, E_2, E_3]$, which represent the spatio-temporal energy ($E_1$),
and the 2-D ($E_2$) and 3-D fractal thermal energies $(E_3$). 
Additional forms of energy definitions in solar and stellar flares, 
such as magnetic energies, radiative energies, conductive energies, 
coronal mass ejection kinetic or potential energies, etc.) 
are studied elsewhere (e.g., Aschwanden et al.~2017). The fact that we can
predict energy size distribution functions, 
[$N_{E_1}$, $N_{E_2}$, $N_{E_3}$], within the statistical uncertainties,
corroborates the validity of the unified scaling laws derived here. 
Specifically, the scaling laws used here involve fractality, 
diffusive transport, flux-volume proportionality, spatio-temporal energy, 
and thermal energies in a fractal volume. The unified
formalism to calculate size distributions based on the scale-free
probability conjecture (Eq.~1) appears to be a sound method to
obtain (macroscopic) physical scaling laws in (microscopic) SOC systems. 
We mention as a caveat however, that 
careful treatment has to be applied to small number statistics, 
truncation biases, data undersampling, background subtraction, 
inadequate fitting ranges, and deviations from ideal power law functions.

\subsection{	Power Law Slopes	}

Our unified method of implementing physical scaling laws in the
calculation of size (or occurrence rate) distribution functions
yields a power law slope $\alpha_x$ for every SOC parameter $x$. Thus
we have a unique correspondence of a scaling law with the power law
slope $\alpha$. Our results yield a power law slope of
$\alpha_{E_1}=(13/9)=1.44$ for a SOC system with spatio-temporal
avalanche energies, a slope of 
$\alpha_{E_2}=(7/3)=2.33$ for the thermal energy in a SOC system with 2-D geometry, and
$\alpha_{E_3}=(9/5)=1.80$ for the thermal energy in a SOC system with 3-D geometry.
We can discard the model with the unphysical fractal 2-D geometry, 
but it explains why some researchers
found relatively high values of $\alpha_E > 2$. So we are
left with relatively low values of $\alpha_E < 2$ for realistic
energy models, such as $\alpha_{E_1} \approx 1.44$ for 
spatio-temporal avalanches, or $\alpha_{E_3} \approx 1.80$
for 3-D fractal avalanches. Although we obtain a well-defined 
value for the power law slope $\alpha_E$ for each size
distribution, we should keep in mind that the estimation
of fractal dimensions has some uncertainties within the
fractal domains, such as in the range of $1 \le D_A \le 2$, and 
$2 \le D_V \le 3$, respectively (Aschwanden and Aschwanden 2008a, 2008b).
In principle, one can 
measure the values of the fractal dimensions $D_A$ and $D_V$ 
from the observed (fitted) power law slopes $\alpha_A$ and $\alpha_V$, 
i.e., $D_A=2/(\alpha_A-1)$, and $D_V=2/(\alpha_V-1)$ (Table 1). 

\subsection{	Nanoflares and Coronal Heating 	}	

It was pointed out early on that powerlaw distributions
$N(E) \propto E^{-\alpha}$ of energies, with a slope flater than the
critical value of
$\alpha_E=2$ imply that the energy integral diverges at the upper end
$E_{max}$, and thus the total energy of the distribution is dominated 
by the largest events (Hudson 1991),
\begin{equation}
        E_{tot} = \int_{E_{min}}^{E_{max}} E * N(E) dE
          = \int_{E_{min}}^{E_{max}} (\alpha -1) E^{1-\alpha_E} dE
          = \left( {\alpha - 1 \over 2-\alpha} \right)
            \left[ E_{max}^{2-\alpha} - E_{min}^{2-\alpha_e} \right] \ .
\end{equation}
In the opposite case, however, when
the powerlaw distribution is steeper than the critical value, it will
diverge at the lower end $E_{min}$, and thus the total energy budget will
be dominated by the smallest detected events, an argument that was
used for dominant nanoflare heating 
(Krucker and Benz 1998). However, subsequent simulations demonstrated
that there exists a strong bias towards a steeper slope 
($\alpha_{E_2}\approx 2.3-2.6$) if the assumption of a constant
line-of-sight depth is assumed ($h_0$ = const), while the application 
of an isotropic geometry ($h=A^{1/2}$) lowers the power law slope to
$\alpha \approx 2.0$ (Parnell and Jupp 2000; Benz and Krucker 2002). 
In our analytical 2-D fractal model we predict a power law slope of 
$\alpha_{E2}=(7/3)\approx 2.33$ (Table 3), 
which agrees well with the spread of observed
values, $\alpha_{E_2}=2.38\pm0.09$ (Table 3). This result shows clearly
that the size distribution of nanoflares has a power law slope
of $\alpha < 2$, for both the spatio-temporal model 
($\alpha_{E_1}=(13/9)\approx 1.44$),
as well as for the 3-D fractal thermal energy model 
($\alpha_{E_3}=(9/5)=1.80$), which implies that the energy
in nanoflares does not diverge at the lower end, $E_{min} \lapprox
10^{24}$ erg, and that nanoflares are not the dominant contributor
to the heating of the solar corona.

\section{	Conclusions 	}

In this study we test whether the {\sl standard self-organized
criticality model} can predict the size (or occurrence frequency)
distribution functions $N(x) dx \propto x^{-\alpha_x}$ of physical
parameters $x$ in solar flares, down to the nanoflare regime
with energies of $E \gapprox 10^{24}$ erg. We focus mostly on energy
parameters, such as the spatio-temporal avalanche energy ($E_1$),
the 2-D fractal energy model ($E_2$), and the more realistic
3-D fractal energy model ($E_3$). For this three energy models, 
power law slopes of $\alpha_{E1}=1.44$, $\alpha_{E2}=2.33$, 
and $\alpha_{E3}=1.80$ are predicted.
We test these predictions from literature values and find
mean slopes of $\alpha_{E1}=1.47\pm0.07$ from 9 data sets (Table 2), 
$\alpha_{E2}=2.38\pm0.09$ from 4 data sets (Table 3), and 
$\alpha_{E3}=1.80\pm0.18$ from 17 data sets (Table 4), which all
are fully self-consistent with the predicted values. 

The related observations
include solar flares observed in hard X-rays, soft X-rays,
and EUV wavelengths, from large flares with energies of
$E \lapprox 10^{33}$ erg down to nanoflares (specified as
EUV transients, coronal brightenings, or blinkers). 
We consider both the spatio-temporal (or standard SOC) model 
as well as the 3-D fractal energy model, based on emission
measure analysis, as realistic tools to quantify the energy 
of flares and nanoflares, while the 2-D version of the
fractal energy model ($E_2$) significantly
over-estimates the power law slope of the energy size distributions.
The analytical approach clearly demonstrates that the size 
distribution of nanoflares has a power law slope of $\alpha < 2$, 
and thus the energy in nanoflares does not diverge at the lower 
end of the size distribuitions, so that nanoflares do not qualify
to be dominant contributors to the heating of the solar corona.

While numerical Monte Carlo-type simulations leave the
option of a super-critical value of $\alpha_E \gapprox 2$ open
(Krucker and Benz 1998; Parnell and Jupp 2000), we demonstrate
in this Letter that this conclusion is true only for the
unrealistic 2-D fractal energy model $E_2$,  
observationally ($\alpha_{E_2}=2.38\pm0.09$), as well as
theoretically ($\alpha_{E_2}=(7/3)\approx 2.33$).
Consequently, the power law slope is flater ($\alpha_E < 2$)
for at least two energy models (the spatio-temporal standard
SOC model $\alpha_{E_1}=1.44$, and the 3-D fractal energy
model $\alpha_{E_3}=1.80$), which implies that 
heating of the corona in active regions 
is dominated by large (M- and X-class) flares.
The same argument holds for Quiet Sun regions, where the
largest events in each size distribution (of nanoflares, 
microflares, EUV transients, coronal brightenings, blinkers, etc.) 
dominate the energy budget (see power law slopes $\alpha_{E_3}$ of 
energies in Table 4), rather than the smallest events.  

\def\ref#1{\par\noindent\hangindent1cm {#1}}

\section*{	References	}

\ref{Aschwanden, M.J. and Parnell, C.E. 2002,
        {\sl Nanoflare statistics from first principles: fractal geometry
        and temperature synthesis}, Astrophys. J. 572, 1048}
\ref{Aschwanden, M.J. and Aschwanden P.D. 2008a,
        {\sl Solar flare geometries: I. The area fractal dimension},
        Astrophys. J. 574, 530}
\ref{Aschwanden, M.J. and Aschwanden P.D. 2008b,
        {\sl Solar flare geometries: II. The Volume fractal dimension},
        Astrophys. J. 574, 544}
\ref{Aschwanden, M.J. 2011,
        {\sl Self-Organized Criticality in Astrophysics. The Statistics
        of Nonlinear Processes in the Universe}, ISBN 978-3-642-15000-5,
        Springer-Praxis: New York, 416p.}
\ref{Aschwanden, M.J. 2012,
        {\sl A statistical fractal-diffusive avalanche model of a
        slowly-driven self-organized criticality system},
        A\&A 539, A2, (15 p)}
\ref{Aschwanden, M.J. 2014,
        {\sl A macroscopic description of self-organized systems and
        astrophysical applications}, ApJ 782, 54}
\ref{Aschwanden, M.J., Caspi, A., Cohen, C.M.S., Holman, G.D., Jing, J., 
	Kretzschmar, M., Kontar, E.P., McTiernan, J.M., O'Flannagain, A., 
	Richardson, I.G., Ryan, D., Warren, H.P., Xu, Y. 2017,
 	{\sl Global energetics of solar flares: V. Energy closure},
 	ApJ 836:17 (17pp)}
\ref{Aschwanden, M.J. 2022,
	{\sl The fractality and size distributions of astrophysical
	slef-organized criticality systems},
	ApJ (subm.}
\ref{Bak, P., Tang, C., and Wiesenfeld, K. 1987,
        {\sl Self-organized criticality: An explanation of 1/f noise},
        Physical Review Lett. 59(27), 381}
\ref{Bak, P., Tang, C., and Wiesenfeld, K. 1988,
        {\sl Self-organized criticality},
        Physical Rev. A {\bf 38}(1), 364}
\ref{Benz, A.O. and Krucker, S. 2002,
        {\sl Energy distribution of microevents in the quiet solar corona},
        ApJ 568, 413}
\ref{Berghmans, D., Clette, F., and Moses, D. 1998,
        {\sl Quiet Sun EUV transient brightenings and turbulence. A panoramic
        view by EIT on board SOHO},
        A\&A 336, 1039}
\ref{Berghmans,D. and Clette,F. 1999,
        {\sl Active region EUV transient brightenings - First Results by
        EIT of SOHO JOP80},
        Solar Phys. 186, 207}
\ref{Brkovic, A., Solanki, S.K., and Ruedi, I. 2001,
        {\sl Analysis of blinkers and EUV brightenings in the quiet Sun
        observed with CDS},
        A\&A 373, 1056}
\ref{Crosby, N.B., Aschwanden, M.J., and Dennis, B.R. 1993,
        {\sl Frequency distributions and correlations of solar X-ray
        flare parameters}, Solar Phys. 143, 275}
\ref{Drake, J.F. 1971,
        {\sl Characteristics of soft solar X-ray bursts},
        Solar Phys. 16, 152}
\ref{Hudson, H.S. 1991,
        {\sl Solar flares, microflares, nanoflares, and coronal heating},
        Solar Phys. 133, 357}
\ref{Joulin, V., Buchlin, E., Solomon, J., and Guennou, C. 2016,
 	{\sl Energetic characterisation and statistics of solar coronal 
	brightenings},
 	A\&A 591, A148}
\ref{Kawai, T. and Imada, S. 2022,
 	{\sl Factors that determine the power-law index of an energy
	distribution of solar flares},
 	ApJ ... (in press)}
\ref{Krucker, S. and Benz, A.O. 1998,
        {\sl Energy distribution of heating processes in the quiet solar
        corona}, Astrophys. J. 501, L213}
\ref{Lu, E.T., Hamilton, R.J., McTiernan, J.M., and Bromund, K.R. 1993,
        {\sl Solar flares and avalanches in driven dissipative systems},
        ApJ 412, 841}
\ref{Nhalil, N.V., Nelson, C.J., Mathioudakis, M., and Doyle, G.J. 2020,
 	{\sl Power-law energy distributions of small-scale impulsive events 
	on the active Sun: results from IRIS},
 	MNRAS 499, 1385}
\ref{Parnell,C.E. and Jupp,P.E. 2000,
        {\sl Statistical analysis of the energy distribution of nano\-flares
        in the quiet Sun}
        ApJ 529, 554}
\ref{Perez Enriquez, R., Miroshnichenko, L.I. 1999,
        Frequency distributions of solar gamma ray events related
        and not related with SPEs 1989 -- 1995,
        Solar Phys., 188, 169}
\ref{Pruessner, G. 2012, {\sl Self-Organised Criticality. Theory, Models
        and Characterisation}, Cambridge University Press: Cambridge.}
\ref{Purkhart S. and Veronig, A.M. 2022,
	{\sl Nanoflare distributions over solar cycle 24 based on SDO/AIA
	differential emission measure observations},
	A\&A (in press).}
\ref{Rosner,R., and Vaiana,G.S. 1978, 
	{\sl Cosmic flare transients: constraints upon models for energy 
	storage and release derived from the event frequency distribution},
	ApJ 222, 1104}
\ref{Shimizu, T. 1995,
        {\sl Energetics and occurrence rate of active-region transient
        brightenings and implications for the heating of the active-region
        corona}, Publ. Astron. Soc. Japan 47, 251.}
\ref{Uritsky, V.M., Paczuski, M., Davila, J.M., and Jones, S.I. 2007,
        {\sl Coexistence of self-organized criticality and intermittent
        turbulence in the solar corona},
        Phys.~Rev.~Lett. 99(2), id. 025001}
\ref{Uritsky, V.M., Davila, J.M., Ofman, L., and Coyner, A.J. 2013,
        {\sl Stochastic coupling of solar photosphere and corona},
        ApJ 769, 62}

\clearpage

\begin{table}
\begin{center}
\normalsize
\caption{Parameters of the standard SOC Model, with fractal dimensions $D_x$
and power law slopes $\alpha_x$ of size distributions).}
\medskip
\begin{tabular}{lll}
\hline
Parameter                    &Power law                                   & Power law  \\
                             &slope                                       & slope      \\
                             &analytical                                  & numerical  \\
\hline
\hline
Euclidean Dimension          &$d =$                                       &3.00        \\
Diffusion type               &$\beta=$                                    &1.00        \\
\hline
Area fractal dimension       &$D_A=d-(3/2)=$                              &1.50=(3/2)  \\
Volume fractal dimension     &$D_V=d-(1/2)=$                              &1.20=(5/2)  \\
\hline
Length                       &$\alpha_L=d=$                               &3.00        \\
Area                         &$\alpha_A=1+(d-1)/D_A=$                     &2.33=(7/3)  \\
Volume                       &$\alpha_V=1+(d-1)/D_V=$                     &1.80=(9/5)  \\
Duration                     &$\alpha_T=1+(d-1) \beta/2=$                 &2.00        \\
Mean flux                    &$\alpha_F=1+(d-1)/(\gamma D_V)=$            &1.80=(9/5)  \\
Peak flux                    &$\alpha_P=1+(d-1)/(\gamma d)=$              &1.67=(5/3)  \\
Spatio-temporal energy       &$\alpha_{E_1}=1+(d-1)/(\gamma D_V+2/\beta)=$ &1.44=(13/9)\\
Thermal energy (h=const)     &$\alpha_{E_2}=1+2/D_A=$                      &2.33=(7/3) \\
Thermal energy (h=A$^{1/2}$) &$\alpha_{E_3}=1+2/D_V=$                      &1.80=(9/5) \\
\hline
\end{tabular}
\end{center}
\end{table}

\begin{table}
\begin{center}
\caption{Observed frequency distributions of spatio-temporal energies $E = F*T$,
by integrating the flux rate $F$ in space and time $T$.}
\medskip
\begin{tabular}{llll}
\hline
Powerlaw slope & Instrument	    & Observed		& Reference 	      \\
of energy      & Wavelength         & Phenomenon	&		      \\
$\alpha_E$     &		    &			&		      \\
\hline
\hline
1.53$\pm$0.02  & HXRBS($>$25 keV)   & solar flares 	& Crosby \etal (1993) \\
1.51$\pm$0.04  & HXRBS($>$25 keV)   & solar flares	& Crosby \etal (1993) \\
1.48$\pm$0.02  & HXRBS($>$25 keV)   & solar flares	& Crosby \etal (1993) \\
1.53$\pm$0.02  & HXRBS($>$25 keV)   & solar flares 	& Crosby \etal (1993) \\
1.51           & ISEE3($>$25 keV)   & solar flares	& Lu \etal (1993)     \\
1.39$\pm$0.01  & PHEBUS($>$100 keV) & solar flares	& Perez-Enriquez and Miroshnichenko (1999)\\
1.44           & Explorer SXR 2-12 A& solar flares	& Drake (1971)	      \\
1.34$\pm$0.08     & SMM/FCS, OV     & blinkers          & Brkovic \etal (2001)\\
1.50$\pm$0.04  & SOHO/EIT 195,HMI   & EUVE events       & Uritsky \etal (2013)\\
\hline
{\bf 1.47$\pm$0.07}&		    &                   & Mean of 9 observations \\
{\bf 1.44}     &		    &      		& Theoretical prediction\\
\hline
\end{tabular}
\end{center}
\end{table}

\begin{table}
\begin{center}
\normalsize
\caption{Observed frequency distributions of thermal energies $E_2$
calculated from peak emission measures and temperatures with 
2-D fractal model and constant line-of-sight depth ($h_0$=const).}
\medskip
\begin{tabular}{llll}
\hline
Powerlaw slope of   	& Instrument    & Observed   	& Reference\\
fluence or energy       & Wavelength    & Phenomenon 	& \\
$\alpha_{E_2}$          &               &            	& \\
\hline
\hline
2.45$\pm$0.15           & EIT 171,195   & EUV transient & Krucker \& Benz (1998)\\
2.30$\pm$0.30           & TRACE 171,195 & Nanoflares    & Parnell \& Jupp (2000)\\
2.48$\pm$0.11           & TRACE 171,195 & Nanoflares    & Parnell \& Jupp (2000)\\
2.31                    & EIT 171,195   & EUV transient & Benz \& Krucker (2002)\\
\hline
{\bf 2.38$\pm$0.09}     &               &               & Mean of 4 observations \\
{\bf 2.33}              &  		&               & Theoretical prediction \\
\hline
\end{tabular}
\end{center}
\end{table}

\begin{table}
\begin{center}
\normalsize
\caption{Observed frequency distributions of thermal energies $E_3$ based on 
3-D fractal model with isotropic line-of-sight depth $h=\sqrt{A}$.}
\medskip
\begin{tabular}{llll}
\hline
Powerlaw slope of & Instrument       & Observed   	    & References\\
fluence or energy & Wavelength       & Phenomenon 	    & \\
$\alpha_{E_3}$    &                  &            	    & \\
\hline
\hline
1.55$\pm$0.05     & Yohkoh   	     & Solar flares	 & Shimizu (1995)      \\
1.90              & SOHO/EIT 195     & EUV transient	 & Berghmans \etal (1998)\\
1.73$\pm$0.28     & SOHO/EIT 195     & EUV transient	 & Berghmans and Clette (1999)\\
2.05$\pm$0.05     & TRACE 171,195    & nanoflares        & Parnell and Jupp (2000)\\
1.57$\pm$0.05     & Yohkoh SXT/AlMg  & nanoflares        & Aschwanden and Parnell (2002)\\
1.41$\pm$0.09     & Yokhoh SXT/AlMg  & nanoflares	 & Aschwanden and Parnell (2002)\\
1.81$\pm$0.10     & TRACE 195        & nanoflares	 & Aschwanden and Parnell (2002)\\
1.70$\pm$0.17     & TRACE 195        & nanoflares	 & Aschwanden and Parnell (2002)\\
1.86$\pm$0.07     & TRACE 171        & nanoflares	 & Aschwanden and Parnell (2002)\\
2.06$\pm$0.10     & TRACE 171        & nanoflares	 & Aschwanden and Parnell (2002)\\
1.66              & SOHO/EIT 195     & nanoflares        & Uritsky \etal (2007)\\
1.79$\pm$0.01     & AIA/SDO 171 A    & coronal brightenings & Joulin \etal (2016) \\
1.83$\pm$0.01     & AIA/SDO 193 A    & coronal brightenings & Joulin \etal (2016) \\
1.88$\pm$0.01     & AIA/SDO 211 A    & coronal brightenings & Joulin \etal (2016) \\
1.80$\pm$0.01     & IRIS             & nanoflares        & Nhalil \etal (2020) \\ 
2.07$\pm$0.02     & IRIS             & nanoflares        & Nhalil \etal (2020) \\ 
2.00$\pm$0.20     & AIA/SDO          & flares  		 & Kawai and Imada (2022) \\
Outliers:	  &		     &                   &                        \\
(2.15$\pm$0.01)$^a$ & AIA/SDO 131 A  & coronal brightenings & Joulin \etal (2016) \\
(2.53$\pm$0.01)$^a$ & AIA/SDO 335 A  & coronal brightenings & Joulin \etal (2016) \\
(2.28$\pm$0.03)$^b$ & AIA/SDO        & nanoflares     	 & Purkhart and Veronig (2022)\\
\hline
{\bf 1.80$\pm$0.18}&                 &                   & Mean of 17 observations \\
{\bf 1.80}        &  	             &                   & Theoretical prediction \\
\hline
\end{tabular}
\end{center}
$(^a$ No large events are detected in the 131 and 335 \ang\ high-temperature bands
	during the time of observations, which causes a steeper power law slope
	(Joulin et al.~2016).
$(^b$ High-energy events could have significant uncertainties since they may heavily
	depend on accurate event combinations between many pixels, one of the most
	challenging steps in the event detection algorithm (Purkhart and Veronig 2022). 
\end{table}

\end{document}